\documentclass[twocolumn,showpacs,amsmath,amssymb]{revtex4-1}
\usepackage{graphicx}
\begin{document}
\title{\Large \bf The General Type N Solution of New Massive Gravity}
\author{\large Haji Ahmedov and Alikram N. Aliev}
\address{Feza G\"ursey Institute, \c Cengelk\" oy, 34684   Istanbul, Turkey}
\date{\today}

\begin{abstract}

We find the most general algebraic type N solution with non-vanishing scalar curvature, which comprises all type N solutions of new massive gravity  in three dimensions. We also give the special forms of this solution, which correspond to certain critical values of the topological mass. Finally, we show that at the special limit, the null Killing isometry of the spacetime is restored and the solution  describes AdS pp-waves.

\end{abstract}

\pacs{04.60.Kz, 11.15.Wx}

\maketitle

\section{Introduction}

Recently,  Bergshoeff, Hohm and  Townsend  formulated a new dynamical theory of massive gravity in three dimensions \cite{bht}. The theory is now called {\it new massive gravity} (NMG) and is described by  the  Einstein-Hilbert (EH) action  complemented  with  a particular higher-derivative correction term. The latter provides  propagating degrees of freedom in the theory, thereby curing dynamically barren property of general relativity in three dimensions \cite{djh}. In contrast to topologically  massive gravity (TMG), formulated a long ago by Deser, Jackiw and Templeton  \cite{djt, deser}, the theory of NMG preserves parity. The associated wave equation contains  fourth-order derivatives of metric perturbations, representing a physical massive graviton with two polarization states. In TMG , the addition  of the higher-derivative Chern-Simons term to  the EH action violates parity, but it also makes the  theory  dynamical with  a single propagating  massive mode. On the other hand, there  exist some similarities  between TMG and NMG theories. First of all, the linearized behavior of the latter in the Minkowski background is similar to that of TMG, resulting in  a unitary theory of two propagating massive graviton modes as long as one uses the reverse sign for the EH term in the action \cite{bht, deser1}. Attempts to extend the NMG theory to all higher dimensions have revealed that only the three-dimensional model is unitary in the tree level \cite{naka, tekin}.

It is unfortunate that unitarity of  both TMG  and NMG theories requires to reverse the usual sign of the EH term in the total action. For instance, due to this  pecularity the mass of  the BTZ black  holes \cite{btz} becomes negative  that in turn makes unsatisfactory the quantum description in context of the AdS/CFT  correspondence. In the case of TMG, significant progress on  this route was achieved  recently in \cite{strom1}. It was shown that at a ``chiral" point, determined by a certain critical value of   the topological mass, the bulk gravitons disappear and the BTZ black holes  have nonnegative masses. That is,  one obtains a unitary chiral quantum theory of gravity with the usual sign of the EH term, which is a  dual of two-dimensional conformal field theory (CFT$_2$) on the boundary. This remarkable result has renewed the interest in TMG   with  the hope of finding other stable vacua for a consistent formulation of quantum gravity in three dimensions \cite{cdww, strom2, chow1, chow2}. In particular, the authors of \cite{chow1} performed the Petrov-Segre type algebraic classification of exact solutions to TMG, showing that almost all existing in the literature homogeneous space solutions locally reduce to either type D, biaxially squashed AdS$_3$   solutions \cite{nutku, gurses, clement1} or type N, AdS pp-waves solutions \cite{nb, oz1, oz2, cdww, gibbons}.  New Kundt type solutions of TMG were found in \cite{chow2}.

A similar  wave of activity has also appeared  in NMG \cite{clement, giri, berg, troncoso, mg, gho}. It was found that NMG admits AdS$_3$ (BTZ), warped AdS$_3$ black holes \cite{bht, clement} as well as   AdS-wave  solutions \cite{giri}. A new class of asymptotically AdS$_3$  black hole solutions to NMG  with  special values of the cosmological term has been discussed in \cite{berg, troncoso}. Some special  cases of Kundt spacetimes and homogeneous space solutions to NMG were also considered in \cite{moha, baka}, respectively. In the quantum context, contrary to the case of TMG, the reconciliation of incompatibility of bulk/boundary theories, in the sense of their unitarity, still remains unsatisfactory, though a number of critical relations between the AdS$_3$ radius and the mass scale do exist as well \cite{berg, liu1,liu2, sinha}. This fact motivates one to look for all possible exact solutions of NMG that could provide stable vacua for satisfactory quantum aspects of the theory.

In a recent paper \cite{ah1},  we began an exhaustive programme  for studying  exact solutions to NMG. In particular, we  found a simple framework that provided  mapping  all  known Petrov-Segre types D and N  exact solutions of TMG into  NMG. Meanwhile, it should be emphasized that  TMG is a strongly constrained theory as it does not admit static solutions besides ``trivial" Einstein solutions \cite{an, df}. However,  NMG  is a much richer theory that  admits solutions which are absent in TMG \cite{berg, ah1}.

In this Letter, we continue the programme of \cite{ah1} and find the most general solution that describes all algebraic type N spacetimes of  NMG. In Sec.II we briefly recall the field equations of cosmological NMG in terms of  a first-order differential operator (resembling a Dirac type operator)
acting on the traceless Ricci tensor.  In Sec.III we introduce a triad  basis consisting of two null and one spacelike vectors and show that for type N geometries, the covariant derivatives of these vectors are  determined only by three scalar functions. In Sec.IV we  derive  the general form  of the spacetime metric with a single unknown function of two variables, which obeys a second-order linear differential equation. In Sec.V we  obtain the most general  algebraic type N solution and discuss its some special limits of interest.

\section{The Field Equations}

The field equations of NMG  were obtained in \cite{bht}  by varying  the action
\begin{eqnarray}
S &=& \frac{1}{16\pi G} \int d^3x \sqrt{-g}\left(R - 2 \lambda -\frac{1}{m^2}\, K \right)\,,
\label{nmgaction}
\end{eqnarray}
with respect to  the spacetime metric. Here $ R= g^{\mu\nu} R_{\mu\nu} $ is the Ricci scalar,  $ \lambda $ is a cosmological term, $ m $  is a mass parameter and
\begin{eqnarray}
K &=& R_{\mu\nu}R^{\mu\nu}- \frac{3}{8} \,R^2\,.
\label{k}
\end{eqnarray}

In a recent paper \cite{ah1}, it was shown  that if one defines
a first-order  differential operator as
 \begin{eqnarray}
{D\hskip -.25truecm \slash}\,\Phi_{\mu\nu} &= & \frac{1}{2}\left( {\epsilon_{\mu}}^{\alpha\beta}\nabla_{\beta} \Phi_{\nu\alpha} + {\epsilon_{\nu}}^{\alpha\beta}\nabla_{\beta} \Phi_{\mu\alpha} \right)\,,
\label{doper1}
\end{eqnarray}
where $ \Phi_{\mu \nu} $ is a symmetric tensor, then its  action on the traceless Ricci  tensor
\begin{eqnarray}
S_{\mu\nu}= R_{\mu\nu}-\frac{1}{3}\,g_{\mu\nu} R\,,
\label{trlessricci}
\end{eqnarray}
yields
\begin{eqnarray}
{D\hskip -.25truecm \slash\, } S_{\mu\nu}&= & - C_{\mu\nu}\,.
\label{Doper}
\end{eqnarray}
Here $ C_{\mu\nu}$ is  the symmetric, traceless and covariantly  constant Cotton tensor  defined as
\begin{eqnarray}
C_{\mu\nu} &= & {\epsilon_{\mu}}^{\alpha\beta}\nabla_{\alpha}\left(R_{\nu\beta} - \frac{1}{4}\, g_{\nu\beta} R\right)\,.
\label{cotton}
\end{eqnarray}
With the  quantities $ {D\hskip -.25truecm \slash}\, $  and  $ S_{\mu\nu} $, the field equations of NMG  with a cosmological term can be put in the form of the massive Klein-Gordon type equation with curvature-squared source term. Thus, one obtains the field equations of NMG in the form
\begin{eqnarray}
\left({D\hskip -.25truecm \slash\,}^2  - m^2 \right) S_{\mu\nu} & = & T_{\mu\nu}\,,
\label{nmgfieldeqs2}
\end{eqnarray}
where the traceless source term  is given  by
\begin{eqnarray}
T_{\mu\nu}& = &  S_{\mu\rho}S^{\rho}_{\,\nu}- \frac{R}{12}\,S_{\mu\nu}- \frac{1}{3}\,g_{\mu\nu} S_{\alpha\beta}S^{\alpha\beta}\,.
\label{source}
\end{eqnarray}
This equation is  also accompanied  by the equation
\begin{eqnarray}
S_{\mu\nu}S^{\mu\nu} + m^2 R - \frac{R^2}{24} & = &  6 m^2 \lambda\,,
\label{newtrfeqs}
\end{eqnarray}
which  is an analogue of the trace equation in \cite{bht}. For a source tensor given by the  relation
\begin{eqnarray}
T_{\mu\nu} & = & \kappa  S_{\mu\nu}\,,
\label{geocond}
\end{eqnarray}
where $ \kappa  $ is  a function of the scalar curvature, which  is fulfilled
for algebraic types D and N  spacetimes, instead of equation (\ref{nmgfieldeqs2}), we have
\begin{eqnarray}
{D\hskip -.25truecm \slash\,}{^2}S_{\mu\nu} & = & \mu^2 S_{\mu\nu}\,,
\label{kgfinal0}
\end{eqnarray}
with
\begin{eqnarray}
 \mu^2 & = &  m^2 + \kappa\,.
\label{kgfinal}
\end{eqnarray}
We recall that in this description the field equations of cosmological TMG  acquire the Dirac type form
\begin{eqnarray}
{D\hskip -.25truecm \slash\,} S_{\mu\nu} & = &  \mu S_{\mu\nu}\,.
\label{tmgdirac}
\end{eqnarray}
Here $ \mu $ is  the mass parameter of TMG. Further details of this description of NMG can be found in \cite{ah1}. We note that in (\ref{cotton}) and in  what follows, the Levi-Civita tensor $\epsilon_{\mu\alpha\beta} $  is given by the relation $\epsilon_{\mu\alpha\beta}= \sqrt{-g} \,\varepsilon_{\mu\alpha\beta} $ and we use the convention $\varepsilon_{012}=1 $.

For algebraic types D and N spacetimes  such a description turns out to be very powerful for finding exact solutions to NMG. This has been  demonstrated in \cite{ah1} by mapping all known algebraic  types D and N  exact solutions of TMG into NMG  as  well as presenting new examples of such solutions, which are only inherent in NMG. In the following, we  further demonstrate the advantages of this description and find the most general type N solution to NMG.\\[2mm]

\section{The Type N Geometries}

We begin with introducing  a triad basis of real vectors, $ \{ l_{\mu}\,,  n_{\mu}\,, m_{\mu}\} $ such that
\begin{eqnarray}
l_{\mu}  n^{\mu} &= & 1\,,~~~ m_{\mu}  m^{\mu}=1\,
\label{norms}
\end{eqnarray}
with all other contractions vanishing identically. That is, we have two null vectors  $ l_{\mu} $ and  $ n_{\mu} $  and one spacelike unit vector  $ m_{\mu} $. The latter vector with
\begin{eqnarray}
m_{\mu} &= &\epsilon_{\mu\nu \sigma} l^{\nu} n^{\sigma}
\label{or1}
\end{eqnarray}
provides an orientation of the three-dimensional manifold. Clearly, we also have
\begin{eqnarray}
l_{\mu} &=& \epsilon_{\mu\nu \sigma} m^{\nu} l^{\sigma}\,,~~~~ n_{\mu} = \epsilon_{\mu\nu \sigma} n^{\nu} m^{\sigma}\,.
\label{or2}
\end{eqnarray}
The spacetime metric written in terms of the basis vectors has the form
\begin{eqnarray}
g_{\mu\nu} &= & 2 l_{(\mu} n_{\nu)} + m_{\mu} m_{\nu}\,.
\label{metric}
\end{eqnarray}
We now recall that for type N spacetimes, in the Petrov-Segre classification of three-dimensional spacetimes, the canonical form of the  traceless Ricci tensor is given by
\begin{eqnarray}
S_{\mu\nu}= l_{\mu}l_{\nu}\,,
\label{Nricci1}
\end{eqnarray}
where  $ l_{\mu} $ is a null vector (see for instance, \cite{chow1, hall, garcia}). With this expression, equation (\ref{newtrfeqs}) yields
\begin{eqnarray}
\lambda= -\nu^2 \left(1+\frac{\nu^2}{4m^2}\right).
\label{cosparam}
\end{eqnarray}
Here, for a  convenience in the following, we have introduced $ \nu^2= - R/6  $, which defines the case of a negative scalar curvature. Equivalently, one can also define it as  $ \nu= \sqrt{-\Lambda}\, $, where  $ \Lambda $ is the usual cosmological constant. Meanwhile, using equation (\ref{geocond}) we have
\begin{eqnarray}
\mu^2 & = &  m^2 +\frac{\nu^2}{2}\,,
\label{Nkg}
\end{eqnarray}
instead of (\ref{kgfinal}).

Next, from the contracted Bianchi identity, $ S_{\mu\nu}^{~~~;\,\nu}=0\, $, where the semicolon stands for covariant differentiation,  we find the relation
\begin{eqnarray}
l_{\nu}^{~ ; \nu} l_{\mu} + l^{\nu}l_{\mu ; \nu} & = & 0\,.
\label{bianchi}
\end{eqnarray}
This enables one to establish the general representation for
the covariant derivative of the  vector $ l_{\mu} $ in the form
\begin{equation}
l_{\mu ; \nu}  =  \alpha  l_{\mu}  l_{\nu}+\beta  l_{\mu} m_{\nu}  +\gamma \left( 2 m_{\mu} m_{\nu}- l_{\mu} n_{\nu}\right)+\sigma  m_{\mu} l_{\nu}\,,
\label{expan1}
\end{equation}
where the coefficients of the expansion are functions of spacetime  and $ \gamma= l_{\mu}^{~ ; \mu} \,$. Using this expression, with equations (\ref{or1}) and (\ref{or2}) in mind,  one can easily calculate the action of the operator $ {D\hskip -.25truecm \slash\,}  $ on the  traceless Ricci tensor $ S_{\mu\nu} $. We find that
\begin{eqnarray}
{D\hskip -.25truecm \slash\, } S_{\mu\nu}&= & \left(\sigma- 2 \beta\right)l_{\mu} l_{\nu}- 4\gamma\, l_{(\mu} m_{\nu)}\,.
\label{ds1}
\end{eqnarray}
Comparing this equation with that given in (\ref{tmgdirac}), we see that for all type N solutions of TMG the function $ \gamma $ must be zero that, along with (\ref{bianchi}), yields
\begin{eqnarray}
l_{\mu}^{~ ; \mu} & = & 0 \,,~~~~~ l^{\nu}l_{\mu ; \nu}  =  0\,.
\label{ku1}
\end{eqnarray}
That is, the null vector $ l_{\mu } $ forms a congruence of expansion-free null geodesics. In other words, all algebraic type N solutions of TMG are Kundt spacetimes \cite{chow2}.

Next, we shall show that the above statement remains  true for the case of NMG as well. For this purpose, it is convenient to begin by assuming that $ \gamma \neq 0 $. Then, the use of the Lorentz transformation of the basis vectors
\begin{eqnarray}
l_{\mu} \rightarrow  l_{\mu}\,,~~~m_{\mu} \rightarrow  m_{\mu}- f l_{\mu}\,,~~~ n_{\mu} \rightarrow n_{\mu} + f m_{\mu}
-\frac{1}{2}\,f^2 l_{\mu}\,,\nonumber\\
\label{lor1}
\end{eqnarray}
where $ f $ is a real function, enables one to set  $ \sigma = 2\beta $ in the new basis. As  a consequence, we  have
\begin{eqnarray}
l_{\mu ; \nu} & = & \alpha  l_{\mu}  l_{\nu}+\beta \left( l_{\mu} m_{\nu} +2  m_{\mu} l_{\nu}\right) +\gamma \left( 2 m_{\mu} m_{\nu}- l_{\mu} n_{\nu}\right). \nonumber \\
\label{lexpan2}
\end{eqnarray}
The use of this expression, along with the  orthogonality condition $ l_{\mu} m^{\mu}=0 $, yields the equation
\begin{eqnarray}
m_{\mu ; \nu} & = & -2\beta  n_{\mu} l_{\nu}- 2 \gamma \,n_{\mu} m_{\nu} +l_{\mu}  x_{\nu}\,,
\label{dm1}
\end{eqnarray}
where  $ x_{\mu} $ is a real vector. It is also straightforward to show that
\begin{eqnarray}
{D\hskip -.25truecm \slash\, } S_{\mu\nu}&= &- 4\gamma\, l_{(\mu} m_{\nu)}\,.
\label{ds2}
\end{eqnarray}
With equations (\ref{lexpan2}), (\ref{dm1}) and  (\ref{ds2}) we are able to calculate the explicit form of $ {D\hskip -.25truecm \slash\,}{^2} S_{\mu\nu} $, which turns out to involve the  term proportional to $ \left(l^{\mu} \gamma_{;\mu} - \gamma^2 \right) \left(g_{\mu\nu}-3 m_{\mu} m_{\nu}\right) $. Clearly, this term must vanish, as a consequence of the field equations in (\ref{kgfinal0}). Thus, we arrive at the equation
\begin{eqnarray}
 l^{\mu} \gamma_{;\mu} - \gamma^2 &= & 0\,.
 \label{zero1}
\end{eqnarray}
We now use the fact that for any vector $ x_{\mu} $,
\begin{eqnarray}
 x_{\nu ;\mu}^{\ \ \ ; \nu}- {(x_\nu^{\ ;\nu})}_{;\mu}&=& R_{\mu\nu }x^\nu= \left(S_{\mu\nu} + \frac{R}{3} \,g_{\mu\nu}\right) x^\nu,
 \label{id1}
\end{eqnarray}
where we have used equation (\ref{trlessricci}). Replacing here $ x_{\nu} $ by $ l_{\nu} $ and then contracting the result with $ l^{\mu} $, we obtain that
\begin{eqnarray}
 l^{\mu} \gamma_{;\mu} +3 \gamma^2 &= & 0\,.
 \label{zero2}
\end{eqnarray}
Combining now equations (\ref{zero1}) and (\ref{zero2}), we see that $ \gamma=0 $, that contradicts with our initial assumption $ \gamma \neq 0 $, i.e. we again arrive at equations in (\ref{ku1}). Thus, {\it all algebraic  type N solutions of NMG are Kundt spacetimes}.

Next, it  is important to establish for these spacetimes the most suitable representation of the covariant derivatives of the basis vectors. We first note that
\begin{eqnarray}
 l^{\mu} \partial_{\mu}\beta &= & 0\,,~~~~~~ l^{\mu} \partial_{\mu}\sigma  =  0\,,
 \label{constr1}
\end{eqnarray}
where, with  (\ref{ku1}) in mind, the first equation is obtained when using expression (\ref{expan1}) in (\ref{id1}) for $ x_{\nu}= l_{\nu} $ and contracting the result with the basis vector $ m^{\mu} $, whereas the second equation follows from the vanishing divergence of equation (\ref{ds1}).  Equations in (\ref{constr1}) allows us to reduce equation (\ref{expan1}) into the form
 \begin{eqnarray}
l_{\mu ;\nu}= \alpha l_\mu l_\nu +\beta \left(l_\mu  m_\nu - l_\nu m_\mu  \right).
\label{trunexpan}
\end{eqnarray}
by means of  the Lorentz transformation of the triad given by $ l_{\mu} \rightarrow k ^{1/2} l_{\mu}\,,~  n_{\mu} \rightarrow  k ^{-1/2} n_{\mu}\,,~  m_{\mu} \rightarrow \, m_{\mu} $, provided that  $ l^{\mu} \partial_{\mu} k=0 $.  We see that  equation (\ref{trunexpan}) involves only two functions $ \alpha $ and $ \beta $.  However, the canonical form of the traceless Ricci tensor now acquires an extra function $ k $,
\begin{eqnarray}
S_{\mu\nu}= k l_{\mu}l_{\nu}\,,
\label{Nricci2}
\end{eqnarray}
not violating the conditions in (\ref{ku1}). Using the fact that expression (\ref{trunexpan}) remains invariant under the Lorentz transformation (\ref{lor1}), one can choose   the vector $ m^ {\mu} $ to be commuting with the  vector $ l^{\mu} $ in the sense of their Lie bracket,
\begin{eqnarray}
\left[l, m\right]=0\,.
\label{Liebr1}
\end{eqnarray}
This in turn allows us to specify the covariant derivative of the vector  $ m $  in the form
\begin{eqnarray}
m_{\mu ;\nu} &= &  \tau l_\mu l_\nu + \beta \left(l_\mu  n_\nu + l_\nu n_\mu  \right )+ \chi l_\mu m_\nu\,.
\label{mexpan1}
\end{eqnarray}
With equations (\ref{trunexpan}) and (\ref{mexpan1}) one can easily  write out the covariant derivative of $ m^ {\mu} $ as follows
\begin{eqnarray}
n_{\mu ;\nu} &= & -\alpha n_{\mu}l_{\nu} - \beta \left(n_{\mu}m_{\nu}+
m_{\mu}n_{\nu}\right)-\tau m_{\mu}l_{\nu}- \chi m_{\mu} m_{\nu}\,.\nonumber\\
\label{nexpan1}
\end{eqnarray}
We note that the Lie bracket in (\ref{Liebr1}) is preserved with respect to transformation (\ref{lor1}), provided that the  function $ f $ obeys the condition $ l^{\mu} \partial_{\mu} f =  0 $. On the other hand, using equations (\ref{trunexpan}), (\ref{mexpan1}) and (\ref{nexpan1}) successively  in (\ref{id1}), after some manipulations, we also find that $ l^{\mu} \partial_{\mu} \chi =  0 $. With these two conditions in mind, one can discard the function $ \chi $ in (\ref{mexpan1}) and (\ref{nexpan1}) by means of the Lorentz transformation in (\ref{lor1}). Thus, we find that {\it for the type N geometries, the covariant derivatives of the  basis vectors are  determined only by three scalar functions}.

\section{The Construction of the Metric}

It is straightforward to show that the associated determining equations for the scalar functions $ \alpha $, $ \beta$ and  $ \tau $ are obtained from equation (\ref{id1}) by an appropriate using  expressions (\ref{trunexpan}), (\ref{Nricci2}), (\ref{mexpan1}) and (\ref{nexpan1}) in it. As a consequence, we have the set of simple equations
\begin{eqnarray}
\label{lindiff1}
\partial_\rho \tau  &=& -k- 4\beta \tau\,,\\[2mm]
\label{lindiff2}
\partial_ v \tau &=& \partial_\rho \alpha +2\alpha \beta =n^{\mu}\partial_{\mu}\beta \,, \\[2mm]
\label{lindiff3}
\partial_ \rho  \beta &=&  -\partial_v \alpha= \nu^2-\beta^2\,,\\[2mm]
\partial_v  \beta &=& 0\,,
\label{lindiff4}
\end{eqnarray}
which are easily solved. Here we have used the definitions $ l = l^\mu \partial_\mu=\partial_v $ and $ m= m^\mu \partial_\mu=\partial_\rho \,$. From equation (\ref{lindiff3}), we immediately see that for $ \nu^2= \beta^2 $, the function $ \alpha $ is constant along the null vector  $ l $. Thus, the null vector determines  a null Killing vector that can be  seen from (\ref{trunexpan}), discarding  the function $ \alpha $ by a Lorentz transformation. This case corresponds to a general AdS pp-waves solution \cite{giri} (see also below, Sec.V).

For $ \nu^2\neq \beta^2 $,  the  most general solutions to the above  set of equations  are given by
\begin{eqnarray}
\alpha  &=& \left(\nu^2-\beta^2\right) \left[- v +\beta  b(u) +c(u) \right],  \\[2mm]
\tau   &=&  \left(\nu^2-\beta^2\right)^2 \left[ v  b(u) +g( u, \rho) \right],
\label{difsetsol1}
\end{eqnarray}
where $ b(u) $ and $ c(u) $ are arbitrary functions of the  coordinate $ u $, the function  $ g( u, \rho) $ obeys the equation
\begin{eqnarray}
\partial_\rho g+\frac{k}{(\nu^2-\beta^2)^2} &= &
0\,,
\label{geq1}
\end{eqnarray}
and $ \beta $ is determined by equations (\ref{lindiff3}) and (\ref{lindiff4}), which admit the following solutions
\begin{eqnarray}
\beta= \nu \tanh(\nu\rho)\,,
\label{betasol1}
\end{eqnarray}
and
\begin{eqnarray}
\beta= \nu \coth(\nu\rho)\,.
\label{betasol2}
\end{eqnarray}
Here we have used a coordinate transformation  to drop
a redundant function of $ u $.

Next, using the Lie brackets
\begin{eqnarray}
[n,l] &=& \alpha l\,,~~~~[n, m] = 2\beta n +\tau l\,,
\label{liebr2}
\end{eqnarray}
established by means of equations (\ref{trunexpan}), (\ref{mexpan1}) and (\ref{nexpan1})  as well as  equation (\ref{lindiff2}), we  find the following representation for the null vector
\begin{eqnarray}
n &= & n^\mu \partial_\mu= \left(\nu^2-\beta^2\right)\left[ A\partial_v+  b(u) \partial_\rho+ \partial_u\right],
\label{nrep}
\end{eqnarray}
where  $ A $ is given by
\begin{eqnarray}
A=\frac{1}{2} \left[v^2 -  a(u,\rho)\right]- v \left[ \beta  b(u) + c(u)\right],
\label{Afunc}
\end{eqnarray}
and  the function  $ a(u,\rho) $ is determined by the equation
\begin{eqnarray}
\partial_\rho a = 2(\nu^2-\beta^2) g\,.
\label{aeq1}
\end{eqnarray}
Taking once again the derivative of this equation with respect to $ \rho $ and combining the result with equations (\ref{lindiff1}) and (\ref{geq1}), we  obtain the second-order linear inhomogeneous differential equation for $ a(u,\rho)$
\begin{eqnarray}
\partial^2_\rho a + 2 \beta \,\partial_\rho a & = &- \frac{2 k}{\nu^2-\beta^2}\,.
\label{aeq2}
\end{eqnarray}
The associated dual 1-forms
\begin{eqnarray}
l_{\mu} dx^{\mu}  & = & \frac{1}{\nu^2-\beta^2}\, du\,,~~ n_{\mu} dx^{\mu}   =  dv- A du\,,\nonumber \\&&
 m_{\mu} dx^{\mu}   =  d\rho- b(u) du\,,
\label{duals}
\end{eqnarray}
define the metric
\begin{eqnarray}
ds^2 & =  & \frac{2}{\nu^2-\beta^2} du \left( dv - A du\right)
 +\left[d \rho - b(u) du\right]^2,
\label{metric1}
\end{eqnarray}
that by means of the coordinate transformation
\begin{equation}
v \rightarrow  v+\beta b(u)+ c(u)\,,
\label{vchange}
\end{equation}
can also be put in the form
\begin{eqnarray}
ds^2 & =  &  d\rho^2 + \frac{2}{\nu^2-\beta^2}\, du dv  +  \frac{1}{\nu^2-\beta^2}\left[a(u,\rho) - v^2 \right]
    du^2\,.\nonumber\\
\label{metric2}
\end{eqnarray}
In obtaining this expression, we have  used the invariance of equations (\ref{geq1}) and (\ref{aeq1}) with respect to the transformation $ a\rightarrow a+ \beta g_1(u)+g_2(u) $, that removes two redundant functions in the metric. Thus, the most general metric for algebraic type N geometries  is characterized  by  a single unknown function   $ a(u,\rho) $ governed by equation (\ref{aeq2}). Clearly, this metric does not admit a null Killing vector field.

The remaining step is to find the explicit form of this function. In doing so, as it follows from equations (\ref{aeq2}), we first need to know the function $ k $.  For this purpose, let us calculate the action of the operator $ {D\hskip -.25truecm \slash\,}  $ on the tensor in (\ref{Nricci2}). We find that
\begin{eqnarray}
{D\hskip -.25truecm \slash\, } S_{\mu\nu}&= & -z S_{\mu\nu}\,,
\label{kdirac1}
\end{eqnarray}
where
\begin{eqnarray}
z &= & 3\beta+ \partial_\rho \ln k\,.
\label{z}
\end{eqnarray}
Comparing this equation with that in (\ref{tmgdirac}), we see that for $ z=\pm \mu $,  we arrive at TMG theory and the corresponding type N solutions recover those found in \cite{chow2}. These solutions of TMG can be mapped into NMG using the prescription given in \cite{ah1}.

It is easy to see that the action of the operator $ {D\hskip -.25truecm \slash\,} $ on equation (\ref{kdirac1}) yields
\begin{eqnarray}
{D\hskip -.25truecm \slash\, }^2 S_{\mu\nu}&= & \left(\partial_\rho z +z^2\right) S_{\mu\nu}\,,
\label{kdirac2}
\end{eqnarray}
that, with the field equations of NMG given in  (\ref{kgfinal0}), leads to the equation
\begin{eqnarray}
\partial_\rho z  & = & \mu^2-z^2\,.
\label{zeq2}
\end{eqnarray}
The nontrival solutions of this equation are given by
\begin{eqnarray}
z &= &\mu \tanh[\mu\rho + h(u)]\,,
\label{zsol1}
\end{eqnarray}
and
\begin{eqnarray}
z &=& \mu \coth[\mu \rho + h(u)]\,,
\label{zsol2}
\end{eqnarray}
where we keep an arbitrary function $ h(u) $ as one can not gauge out it simultaneously with that entering into the  solutions for $ \beta $ (see equations (\ref{betasol1}) and  (\ref{betasol2})).

Next, combining  equations (\ref{z}) and (\ref{zeq2}) and taking into account equation (\ref{lindiff3}), we find that
\begin{eqnarray}
k &=& -\frac{1}{2}\,(\nu^2-\beta^2)\left(\frac{\nu^2-\beta^2}{\mu^2-z^2}\right)^{1/2} F(u)\,,
\label{ksol}
\end{eqnarray}
where $ F(u) $ is an arbitrary function.

\section{The general solution}

Substituting now  the expression of $ k $ given above into equation (\ref{aeq2}), making use of  solutions  (\ref{betasol1}), (\ref{betasol2}) and  (\ref{zsol1}),  (\ref{zsol2}), we can  solve it for the metric function $ a(u,\rho) $.

We first consider the generic case $ \mu^2 \neq \nu^2 $ and  begin with  $\beta = \nu \tanh(\nu\rho)$ . Then,  for the function $ z $ given in equations (\ref{zsol1}) and  (\ref{zsol2}), the solutions of equation (\ref{aeq2}) are respectively given by
\begin{equation}
a = \frac{\cosh[\mu\rho+h(u)]}{\cosh(\nu\rho)}\,f(u) +\tanh(\nu\rho)\,  f_1(u) + f_2(u),
\label{asol1}
\end{equation}
and
\begin{equation}
a = \frac{\sinh[\mu\rho+h(u)]}{\cosh(\nu\rho)}\,f(u) +\tanh(\nu\rho)\,  f_1(u) + f_2(u).
\label{asol2}
\end{equation}
Since $ h(u) $  is an arbitrary function, these two solutions can be ``glued" together to give the single solution
\begin{eqnarray}
a & =& \frac{1}{\cosh(\nu\rho)}\left[\cosh(\mu\rho) \,F_1(u)+ \sinh(\mu\rho) \,F_2(u)\right. \nonumber \\[2mm]  & & \left.
+ \cosh(\nu\rho) \, f_1(u) + \sinh(\nu\rho)\,
    f_2(u)\right],
\label{asingle1}
\end{eqnarray}
which involves four arbitrary functions of $ u $. We recall that this solution corresponds to the case of negative scalar curvature. However, making  an analytical continuation $ \nu\rightarrow i\nu  $ or taking the limit  $ \nu\rightarrow 0 $ , we obtain the solutions corresponding to the positive or zero values of the scalar curvature, respectively. It is also important to note that, as it follows from equation (\ref{Nkg}), the quantity $ \mu^2 $ can take on both negative and zero values. The associated solutions are also recovered by (\ref{asingle1}) when performing  an analytical continuation $ \mu \rightarrow i\mu   $ or taking the limit  $ \mu \rightarrow 0 $.

Similarly, for  $\beta = \nu \coth(\nu\rho)$ and  $ \mu^2 \neq \nu^2 $, we have the solution
\begin{eqnarray}
a & =& \frac{1}{\sinh(\nu\rho)}\left[\cosh(\mu\rho) \,F_1(u)+ \sinh(\mu\rho) \,F_2(u)\right. \nonumber \\[2mm]  & & \left.
+ \cosh(\nu\rho) \, f_1(u) + \sinh(\nu\rho)\,
    f_2(u)\right].
\label{asingle2}
\end{eqnarray}
As in the previous case, one can make appropriate analytical continuations to include the case of positive or zero scalar curvature as well as  the case of negative or zero $ \mu^2 $.

It is interesting to note that  the most general solution, comprising  all these solutions,  is given by
\begin{equation}
ds^2  =   d\rho^2 + \frac{2 du dv}{\nu^2-\beta^2}+\left[Z(u,\rho) - \frac{v^2}{\nu^2-\beta^2} \right]du^2,
\label{unimetric}
\end{equation}
where the metric function has  the form
\begin{eqnarray}
Z(u,\rho) & = & \frac{1}{\sqrt{\nu^2-\beta^2}}\left[\cosh(\mu\rho)\, F_1(u)+ \sinh(\mu\rho)\, F_2(u)\right. \nonumber \\[2mm]  & & \left.
+ \cosh(\nu\rho)\,  f_1(u) + \sinh(\nu\rho)\,  f_2(u)\right]
\label{zetunimt1}
\end{eqnarray}
and $ \beta $ is given as in either  (\ref{betasol1}) or (\ref{betasol2}). We recall that $\nu $ is related to the cosmological term $ \lambda $ as given in (\ref{cosparam}) and the general metric does not admit a null Killing vector field. For $ F_2(u) =f_1(u)=f_2(u) = 0 $ and $ \beta $  in (\ref{betasol1}),
this solution reduces to that obtained in \cite{ah1}. We see that the general solution involves two extra functions $ f_1(u) $ and  $ f_2(u) $. In fact, one of these functions  can be discarded by means of coordinate transformations. In order to show  this, it is convenient to write equation (\ref{zetunimt1}) in the following alternative form
\begin{eqnarray}
Z(u,\rho) & = & \frac{\cosh(\mu\rho)\, F_1(u)+ \sinh(\mu\rho)\, F_2(u)}{\sqrt{\nu^2-\beta^2}}
\nonumber \\&&
+ \,\frac{h_1(u) + \beta h_2(u)}{\nu^2-\beta^2}\,.
\label{zetunimt2}
\end{eqnarray}
Here $ h_1(u)\rightarrow f_1(u) $ and  $ h_2(u)\rightarrow f_2(u) $
for $ \beta $ given in (\ref{betasol1}), whereas  $ h_1(u)\rightarrow f_2(u) $  and $ h_2(u)\rightarrow f_1(u) $  for $ \beta $ given in (\ref{betasol2}). Passing now to the  new coordinates $ v\rightarrow  v\, G(u) + H(u) $, such that  $  \partial_u G=  G(u) H(u) $, and  $ du \rightarrow  du/G(u) $, it is  straightforward to see that one can eliminate  $ h_1(u) $ by choosing $ G(u) $ and $ H(u) $. As a result, we have
\begin{eqnarray}
Z(u,\rho) & = & \frac{\cosh(\mu\rho)\, F_1(u)+ \sinh(\mu\rho)\, F_2(u)}{\sqrt{\nu^2-\beta^2}} \nonumber \\&&
+ \sinh(2\nu\rho) F_3(u)\,.
\label{zetunimt3}
\end{eqnarray}
Thus, the most general  solution is characterized  by three arbitrary functions. In the following, for some future purposes, we keep both functions $ f_1(u) $ and  $ f_2(u) $. We proceed with  the special forms of  (\ref{unimetric}), which are of interest as well.

(i) $ \mu^2 =\nu^2 $ (or, as it follows from (\ref{Nkg}), $ m^2=\nu^2/2 $). In this case, taking properly the limit of (\ref {zetunimt1}) or equivalently solving equation (\ref{aeq2}), with the explicit forms of $ \beta $ and $ z $ given above, we find that
\begin{eqnarray}
Z(u,\rho) & = & \frac{1}{\sqrt{\nu^2-\beta^2}} \left \{\cosh(\mu\rho) \left[\rho F_1(u)+f_1(u)\right] \right. \nonumber \\[2mm]  & & \left.
+ \sinh(\mu\rho) \left[\rho F_2(u)
+f_2(u)\right]\right\}.
\label{zetcritical}
\end{eqnarray}

(ii) $ \mu^2 =0 $  (or,  as it follows from (\ref{Nkg}), $ m^2=-\nu^2/2 $). Then, from equation (\ref{zetunimt1}) it immediately follows that
\begin{eqnarray}
Z(u,\rho) & = & \frac{1}{\sqrt{\nu^2-\beta^2}}\left[F_1(u)+ \rho\, F_2(u)\right. \nonumber \\[2mm]  & & \left.
+ \cosh(\nu\rho)\,  f_1(u) + \sinh(\nu\rho)\,  f_2(u)\right].
\label{zetmu0}
\end{eqnarray}
Again, one can  discard the redundant function $ f_1(u) $ in equations (\ref{zetcritical})  and (\ref{zetmu0}) by means of coordinate transformations.

iii) $ \beta^2 \rightarrow  \nu^2  $. In this case, making the  coordinate transformation $ v\rightarrow (\nu^2-\beta^2)\, v $, one can  put  the metric in (\ref{unimetric})  in the form
\begin{eqnarray}
ds^2 &=& d\rho^2 + 2 du dv  -4\beta  v\, d\rho  du \nonumber\\[2mm] &&
+  \left[Z(u,\rho)
- (\nu^2-\beta^2)v^2\right]du^2.
\label{pp1}
\end{eqnarray}
We see that in the limit $ \beta^2 \rightarrow  \nu^2 $,  the term proportional to  $ v^2 $ disappears. That is, the Killing isometry of the spacetime is restored and the vector $ \partial_v $ becomes a  null Killing vector. This in turn means that the resulting metric must describe AdS pp-waves. Indeed, redefining once again the coordinate $ v $  as $ v\rightarrow e^{2\nu\rho}\,v $,  we arrive at the metric
\begin{eqnarray}
ds^2 &=& d\rho^2 +  2 e^{2\nu\rho} du dv  +\left[e^{\nu\rho }\cosh(\mu\rho)  F_1(u) \right. \nonumber \\[2mm]  & & \left.
+ e^{\nu\rho} \sinh(\mu\rho) F_2(u)+ e^{2\nu\rho}\,f_1(u)+f_2(u)
\right]du^2,\nonumber\\
\label{pp2}
\end{eqnarray}
where, in the contrary with the general case (\ref{unimetric}), both functions  $ f_1(u) $  and $ f_2(u) $  can be gauged out by means of coordinate  transformations. This metric is nothing but the AdS pp-waves solution of NMG  that was earlier  found in \cite{giri}.

iv) Finally, for $ F_1(u)= \pm  F_2(u)= f(u) $, the solution in (\ref{unimetric}) recovers that obtained from the TMG case (see Refs.\cite{chow2, ah1}). The same remains true for solution (\ref{zetcritical}) as well.

Thus,  we have the most general solution given in  (\ref{unimetric}), which  comprises  all algebraic type N spacetimes of NMG.

\section{Conclusion}

A novel  description of NMG in three dimensions, given in our previous work \cite{ah1}, turns out to be a very powerful tool for finding exact solutions to the theory. In this paper, we have further demonstrated the advantages of this formalism. Describing three-dimensional  algebraic type N spacetimes in terms of a triad basis (with two null vectors and one spacelike vector), we have shown that  all such solutions of NMG are Kundt spacetimes. Using this property along with  freedoms, provided by the Lorentz  symmetries underlying the system, we have found that  the covariant  derivatives of the  basis vectors are in general  determined only by three scalar functions, which obey a chain of simple equations.

For algebraic type N geometries, we have obtained the general form  of the spacetime metric with a single unknown function of two variables. Remarkably, this function is governed by a second-order linear inhomogeneous differential equation. Finally, solving the linear differential equation, we have found the most general  algebraic type N solution with non-vanishing scalar curvature, which comprises all type N solutions to the theory of NMG. We have also considered some special cases of interest.


\begin{thebibliography}{99}

\bibitem{bht} E. A. Bergshoeff, O. Hohm and P. K. Townsend, Phys. Rev. Lett. {\bf 102} (2009) 201301.
\bibitem{djh}  S. Deser, R. Jackiw and G. 't Hooft, Ann. Phys. (N.Y.)
{\bf 152} (1984) 220.
\bibitem{djt} S. Deser, R. Jackiw and S. Templeton,
 Phys. Rev. Lett. {\bf 48} (1982) 975; Ann. Phys. (N.Y.)
{\bf 140} (1982) 372; {\it erratum-ibid}. {\bf 185} (1988) 406.
\bibitem{deser} S. Deser, in {\it Quantum Theory of Gravity}, ed. S. Christensen, A. Hilger Ltd, London (1984).
\bibitem{deser1} S. Deser, Phys. Rev. Lett. {\bf 103} (2009) 101302.
\bibitem{naka} M. Nakasone and I. Oda, Prog. Theor. Phys. {\bf 121} (2009) 1389.
\bibitem{tekin} \.{I} G\"{u}ll\"{u} and B. Tekin, Phys. Rev. D {\bf 80} (2009) 064033.
\bibitem{btz}  M. Banados, C. Teitelboim and J. Zanelli,  Phys. Rev. Lett. {\bf 69} (1992) 1849.
\bibitem{strom1}  W. Li, W. Song and A. Strominger, J. High Energy Phys. {\bf 0804} (2008) 082.
\bibitem{cdww}  S. Carlip, S. Deser, A. Waldon and D. K. Wise, Phys. Lett B {\bf 666} (2008) 272.
\bibitem{strom2} D. Anninos, W. Li, M. Padi, Wei Song and  A. Strominger, J. High Energy Phys. {\bf 0903} (2009) 130.
\bibitem{chow1} D. K. Chow, C.N. Pope and E. Sezgin, Class. Quant. Grav. {\bf 27} (2010) 105001.
\bibitem{chow2} D. K. Chow, C.N. Pope and E. Sezgin, Class. Quant. Grav. {\bf 27} (2010) 105002.
\bibitem{nutku} Y. Nutku, Class. Quant. Grav. {\bf 10}  (1993) 2657.
\bibitem{gurses} M. G\"{u}rses, Class. Quant. Grav. {\bf 11} (1994) 2585.
\bibitem{clement1} K. Ait Moussa, G. Cl´ement and C. Leygnac, Class. Quantum Grav. {\bf 20} (2003) L277.
\bibitem{nb} Y. Nutku and P. Baekler, Ann. Phys. (N.Y.)
{\bf 195} (1989) 16.
\bibitem{oz1} T. Dereli and \"{O}. Sar{\i}o\v{g}lu, Phys. Rev. D {\bf 64} (2001) 027501.
\bibitem{oz2} S. \"{O}lmez, \"{O}. Sar{\i}o\v{g}lu and B. Tekin, Class. Quant. Grav. {\bf 22} (2005) 4355.
\bibitem{gibbons} G. W. Gibbons, C.N. Pope and E. Sezgin, Class. Quant. Grav. {\bf 25} (2008) 205005.
\bibitem{clement} G. Clement, Class. Quant. Grav. {\bf 26} (2009) 105015.
\bibitem{giri}  E. Ayon-Beato, G. Giribet and M. Hassaine,  J. High Energy Phys. {\bf 0905} (2009)  029.
\bibitem{berg} E. A. Bergshoeff, O. Hohm and P. K. Townsend, Phys. Rev. D {\bf 79} (2009) 124042.
\bibitem{troncoso}  J. Oliva,  D. Tempo and R. Troncoso, J. High Energy Phys. {\bf 0907} (2009) 011.
\bibitem{mg} M. G\"{u}rses, arXiv:1001.1039 [gr-qc]
\bibitem{gho} A. Ghodsi and  M. Moghadassi,  arXiv:1007.4323 [hep-th]
\bibitem{moha} M. Chakad, arXiv:0907.1973 [hep-th]
\bibitem{baka} I. Bakas and  C. Sourdis,  arXiv:1006.1871 [hep-th]
\bibitem{liu1} Y. Liu and Y. W. Sun, J. High Energy Phys. {\bf 0904} (2009) 106.
\bibitem{liu2} Y. Liu and Y. W. Sun, J. High Energy Phys. {\bf 0905} (2009) 039.
\bibitem{sinha}  A. Sinha, J. High Energy Phys. {\bf 1006} (2010) 061.

\bibitem{ah1} H. Ahmedov and  A. N. Aliev, arXiv:1006.4264 [hep-th]
 \bibitem{an} A. N. Aliev and Y. Nutku, Class. Quant. Grav. {\bf 13} (1996)  L29.
\bibitem{df} S. Deser and  J. Franklin, Class. Quant. Grav. {\bf 27} (2010) 1007002.
\bibitem{hall} G. S. Hall, T.  Morgan and Z. Perjes, Gen. Relat. Grav. {\bf 19} (1987) 1137.
\bibitem{garcia} A. Garc\'{\i}a, F. W. Hehl,  C. Heinicke and A. Mac\'{\i}as,

Class. Quant. Grav. {\bf 21} (2004) 1099.



\end{thebibliography}
\end{document}